\documentclass[useAMS]{mn2e}
\usepackage{psfig}

\usepackage{times}

\def\spose#1{\hbox to 0pt{#1\hss}}
\newcommand\lsim{\mathrel{\spose{\lower 3pt\hbox{$\mathchar"218$}}
     \raise 2.0pt\hbox{$\mathchar"13C$}}}
\newcommand\gsim{\mathrel{\spose{\lower 3pt\hbox{$\mathchar"218$}}
     \raise 2.0pt\hbox{$\mathchar"13E$}}}

\title[Spine--layer interplay and TeV $\gamma$-rays from M87]
{Spine--sheath layer radiative interplay in subparsec--scale jets and
the TeV emission from M87}

\author[Tavecchio \& Ghisellini] {Fabrizio Tavecchio\thanks{E--mail:
fabrizio.tavecchio@brera.inaf.it} and Gabriele Ghisellini\\
INAF/Osservatorio Astronomico di Brera, via E. Bianchi 46, I--23807
Merate, Italy}

\begin{document}


\pagerange{\pageref{firstpage}--\pageref{lastpage}} \pubyear{2007}

\maketitle

\label{firstpage}

\begin{abstract}
Simple one--zone homogeneous synchrotron self--Compton models have
severe difficulties is explaining the TeV emission observed in the
radiogalaxy M87.  Also the site the TeV emission region is uncertain:
it could be the unresolved jet close to the nucleus, analogously to
what proposed for blazars, or an active knot, called HST--1, tens of
parsec away. We explore the possibility that the TeV emission of M87
is produced in the misaligned subpc scale jet. We base our modelling
on a structured jet, with a fast spine surrounded by a slower layer.
In this context the main site responsible for the emission of the TeV
radiation is the layer, while the (debeamed) spine accounts for the
emission from the radio to the GeV band: therefore we expect a more
complex correlation with the TeV component than that expected in
one--zone scenarios, in which both components are produced by the same
region.  Observed from small angles, the spine would dominate the
emission, with an overall Spectral Energy Distribution close to those
of BL Lac objects with a synchrotron peak located at low energy
(LBLs).
\end{abstract}

\section{Introduction}
%
%
An increasing number of extragalactic objects
is detected at energies greater than 100 GeV.
The list of published sources comprises 
20 objects\footnote{see {\tt
http://www.mppmu.mpg.de/$\sim$rwagner/sources}}: as expected (e.g.,
Costamante \& Ghisellini 2002), the bulk of them (17) belongs to the
class of Highly Peaked BL Lac objects. 
In fact, the position of the synchrotron peak, 
usually located in (or close to) the X--ray band
assures the existence, in these sources, of electrons with extremely
large Lorentz factors ($\gamma=10^5$--$10^6$), a key ingredient for
the emission in the TeV band via inverse Compton (IC) scattering. 
The remaining three sources are BL Lac itself, which belongs to the LBL
class, (Albert et al. 2007), 3C279, a FSRQs, (Teshima et al. 2007) and
the nearby (16 Mpc, Tonry 1991) radiogalaxy M87 (Aharonian et
al. 2003, 2006).

Already suggested as a possible source of high-energy radiation
(e.g. Bai \& Lee 2001) based on its similarity with BL Lac objects
(Tsvetanov et al. 1998), M87 was discovered as a TeV source by the
HEGRA array of Cherenkov telescopes (Aharonian et al. 2003) and it has
been recently confirmed by H.E.S.S. (Aharonian et al. 2006). Due to
the limited spatial resolution it is not possible to identify the
emission region of this radiation. However, the sensitivity of
H.E.S.S. allowed the detection of variability on short timescales
($\sim$ days), suggesting a compact emission region.

Three main sites have been proposed as emission regions for the TeV
radiation: the resolved jet (in particular the so--called knot HST--1);
the unresolved base of the jet (in analogy with blazars); the vicinity
of the supermassive ($M=3\times 10^9$ $M_{\odot}$, Marconi et
al. 1997) black hole (BH).

HST--1 is a quite interesting jet feature located at 60 pc (projected)
from the core of of M87. Superluminal motions with apparent speeds up
to $\sim 6 c$ were observed with {\it HST} (Biretta et al. 1999).
Continuous multifrequency coverage has shown spectacular activity from
this knot in the past years (Perlman et al. 2003; Harris et al. 2006,
2007; Cheung et al. 2007). In particular, the X--ray flux increases by
a factor of $\sim 30$ from 2000 to 2005 and since then it is steadily
decreasing. These variations are accompanied by similar changes in the
optical and in the radio bands and by the appearance of new
superluminal components. This extreme phenomenology is described by
Stawarz et al. (2006) on the assumption that HST-1 marks the
recollimation shock of the jet. As such, HST--1 is thought to be a
rather efficient particle accelerator and thus a possible source of
intense TeV radiation. However, the small variability timescales
observed by H.E.S.S. put strong limits to the size of the source 
($R < c \Delta t \, \delta \simeq 5\times 10^{15} \delta$ cm, where 
$\delta=[\Gamma(1-\beta \cos\theta)]^{-1}$
is the relativistic Doppler factor, $\theta$ is the viewing angle 
and
$\Delta t\sim 2$ days), which seems difficult to accomplish in this
scenario (e.g. Neronov \& Aharonian 2007; Levinson 2007).

The rapid variability could be easily reproduced if the TeV emission
is produced close to the base of the jet, in the region associated to
the emission from blazars.  However, standard one--zone leptonic
models face difficulties in reproducing the observed spectral energy
distribution (SED) of the core including the TeV data.  A possible
solution advocates that the emission comes from a decelerating jet
(Georganopoulos et al. 2005) .  Alternatively, the emission can be
produced by high--energy protons (Reimer et al. 2004).  Going further
in towards the BH, TeV photons could be emitted through IC by
relativistic pairs produced by electromagnetic cascades in the BH
magnetosphere (Neronov \& Aharonian 2007).

The determination of the site producing the high-energy emission
is rather important. Indeed, if the emission site will be eventually
identified as knot HST-1, this would have a broad impact on the
current view of relativistic jets (see the discussion in Cheung et
al. 2007). It is thus extremely important to investigate whether the
standard view (or minimal variations of it) for the inner jet is able
reproduce the observed phenomenology. If not, it would be mandatory to
explore the other alternatives.

As discussed in Ghisellini, Tavecchio \& Chiaberge (2005, hereafter
Paper I), there is compelling evidence supporting the view that jets
of BL Lac objects are structured at blazar scales (at distances $\sim
10^{17}$ cm from the BH), with a fast core ({\it spine})
surrounded by a slower sheet ({\it layer}).  In Paper I we showed that
the radiative coupling between the spine and the layer can naturally
account for the deceleration inferred for the jet between the blazar
and the VLBI scale.  Furthermore, one component sees the radiation of
the other boosted, since the relative velocity can be relativistic.
This enhances the IC emission of both the spine and the
layer.

A direct consequence of such a structure is that while at small
viewing angle (as in the case of blazars) the emission is dominated by
the boosted spine emission, at large observing angles
($\theta>45^\circ$, typical for radio--galaxies) the emission from the
spine would be suppressed, while the layer, characterised by a broader
beaming cone, could substantially contribute to (and sometimes
dominate) the overall emission. At intermediate angles both components
can significantly contribute.  Following this line, in this letter we
explore an alternative interpretation for the TeV emission of M87,
assuming that the TeV radiation is produced in the layer of the
misaligned jet.
We use $H_{\rm 0}\rm =70\; km\; s^{-1}\; Mpc^{-1} $,
$\Omega_{\Lambda}=0.7$, $\Omega_{\rm M} = 0.3$.

\section{The spine--layer scenario}

\subsection{Difficulties with the one--zone models}
\label{sec1}

\begin{figure}
\vskip -0.7 true cm
\centerline{ \psfig{file=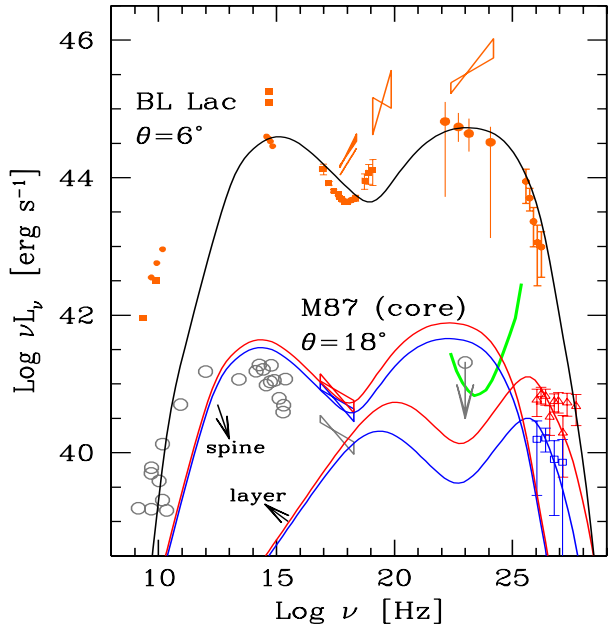,height=10cm,width=12cm} }
\vskip -0.8 true cm
\caption{ SED of the core of M87 (grey open points) together
with the H.E.S.S. spectra taken in 2004 (open blue squares) and
2005 (open red triangles), from Aharonian et al. (2006). The grey
bow--tie reports the X--ray spectrum as measured by {\it Chandra} in
2000 (from Balmaverde et al. 2006).  We reproduced the increased
X--ray emission of 2004 and 2005 assuming the same slope and a larger
normalisation (blue and red bow--ties). The lines report the emission
from the spine and from the layer for the two states as calculated
with the parameters reported in Tab. 1. For comparison the green line
indicates the sensitivity of {\it GLAST} ($5\sigma$, 1 year, converted
in luminosity assuming the distance of M87). The upper black line
reports the emission from the spine that would detect an observer
located at an angle of 6$^\circ$ from the jet axis (the corresponding
emission from the layer is well below that of the spine and is not
reported for semplicity). For comparison, we show the data of BL Lac
(filled orange symbols, from Ravasio et al. 2002, TeV data from Albert
et al. 2007). See text for details.}
\label{fig1}
\end{figure}

The SED of the core of M87 is reported in Fig. \ref{fig1} (grey
open points, see Paper I for references) together with the
H.E.S.S. spectra for 2004 (open blue squares) and 2005 (
open red triangles) as reported by Aharonian et al. (2006).  The
X--ray spectrum (as observed in July 2000) is taken from Balmavarde,
Capetti \& Grandi (2006).  Note that the slope [$\alpha =1.3\pm 0.1$;
$F(\nu)\propto \nu^{-\alpha}$] is intermediate between that reported
in Marshall et al. (2002) ($\alpha =1.47 \pm 0.08$, see also Donato,
Sambruna \& Gliozzi 2004) and that of Wilson \& Yang (2002) ($\alpha
=1.17\pm0.1$).  The X--ray spectral shape was not determined in 2004
and 2005, and for simplicity we assume that it is the same of 2000,
with a normalisation that is a factor 3 (2004, blue) and 4 (2005, red)
greater.  This reproduces the increase of the X--ray flux of the core
recorded through the continuous monitoring of {\it Chandra} (Harris et
al. 2007).

The data describe a bump peaking at $\sim 10^{14}$ Hz and extending
into the X--ray band, while the TeV emission clearly belongs to a
second component. 
The SED closely resembles that usually
observed in blazars, in which the first component is synchrotron
emission from relativistic electrons and the second peak is produced
through IC emission (e.g., Ghisellini et al. 1998). For M87 there is no
evidence of powerful sources of soft photons in the nucleus and hence
the high--energy emission should be dominated by the Synchrotron
self--Compton emission (SSC; e.g., Maraschi et al. 1992). 
However, a simple one--zone synchrotron--SSC model cannot easily describe 
the observed SED. 
The reason is that in order to have the synchrotron
component peaking in the IR region ($\nu_s\simeq 10^{14}$ Hz) and the
SSC one close to the TeV band ($\nu _C\simeq 10^{26}$ Hz, as required
by the relatively flat TeV spectra) one is forced to assume a rather
low magnetic field and an unreasonably large Doppler factor. Following
Tavecchio, Maraschi \& Ghisellini (1998) and defining the synchrotron
and SSC peak luminosities, $L_s\equiv \nu_sL(\nu_s)$ and $L_C\equiv
\nu_CL(\nu_C)$ and the variability timescale $\Delta t$, one obtains:
\begin{equation}
\delta\simeq 500\, L_{s,41.5}^{1/2} L_{C,40.5}^{-1/4} \, \nu_{s,14}^{-1} \, 
\nu_{C,26}^{1/2} \, \Delta t_{1d}^{-1/2}
\end{equation}
Here $Q=10^x Q_x$, in cgs units, except for $\Delta t$, that is in days.
The adopted numerical values are appropriate for M87.

This is a general problem faced by any synchrotron--SSC model using a
single emission region. A direct way to overcome this problem is to
``decouple'' synchrotron and the IC components, assuming that the two
components are produced by two different regions.  Such a
``two-zones'' model would allow us to reproduce the low--frequency
bump (emitted by electrons with relatively low energy) with one
emission component and the TeV emission with a separate source with
high--energy electrons. This opportunity is directly offered by the
spine--layer scenario. In this case the debeamed spine (or,
alternatively, the layer) could be responsible for the low--energy
bump and the layer (or the spine) of the TeV component.

\subsection{The model}
\label{sec2}

We focus on the case of a sub--pc scale emission region, due to the
observed inter--day variability of the TeV emission, difficult to
explain if HST--1 is responsible for this emission.

We are aware that we are almost doubling the number of the free
parameters (one set for the spine, one set for the layer), and
therefore we pay the increased flexibility with an increased
uncertainty concerning the value of the parameters themselves and a
loss of predictive power. On the other hand, our main aim at this
stage is not much to find whether/what specific physical conditions
(value of the magnetic field, beaming factor, number of particles, and
so on) are univocally determined by the model, but to show that one
can attribute the radiation we see to the the sub--pc jet (with
``minor'' modification of the simplest model) without radically
changing our ideas about the emission models of blazars. Having said
that, we will nevertheless do our best to find the most reasonable set
of parameters.  This is why we will not only try to fit the SED of
M87, but we will worry about what kind of blazar would M87 be if seen
at small angles.  If the resulting SED would be too odd with respect
to the known blazar SED, we will discard the found parameter set.

We refer to Paper I for a complete description of the model.  Briefly,
we assume that the jet comprises two regions: i) the spine, assumed to
be a cylinder of radius $R$, height $H_s$ (as measured in the spine
frame) and in motion with bulk Lorentz factor $\Gamma _s$; ii) the
layer, modelled as an hollow cylinder with internal radius $R$,
external radius $R_2$, height $H_l$ (as measured in the frame of the
layer) and bulk Lorentz factor $\Gamma_l$. Each region is
characterised by tangled magnetic field with intensity $B_{s}$,
$B_{l}$ and it is filled by relativistic electrons assumed to follow a
smoothed broken power--law distribution extending from $\gamma_{min}$
to $\gamma_{cut}$ and with indices $n_1$, $n_2$ below and above the
break at $\gamma_b$.  The normalisation of this distribution is
calculated assuming that the system produces an assumed synchrotron
luminosity $L_{syn}$ (as measured in the local frame), which is an
input parameter of the model.  As in Paper I we assume that $H_l >
H_s$.  The seed photons for the IC scattering are not only those
produced locally in the spine (layer), but we also consider the
photons produced in the layer (spine).  Hereafter we will indicate
this component as External Compton (EC).

To properly calculate the observed emission, it is necessary to take
into account the different beaming patterns associated to the two
emission components.  This is because, in the comoving frame of the
spine (layer), the photons produced by the layer (spine) are not
isotropic, but are seen aberrated. Most of them are coming from almost
a single direction (opposite to the relative velocity vector).  In
this case the produced EC is anisotropic even in the comoving frame,
with more power emitted along the opposite direction of the incoming
photons (i.e. for head--on scatterings).  We therefore must take into
account this anisotropy in the comoving frame when transforming in the
observer frame.  Consider also that this applies only to the EC
radiation, while the synchrotron and the SSC emissions are isotropic
in the comoving frame.  The EC radiation pattern of the spine will be
more concentrated along the jet axis (in the forward direction) with
respect to its synchrotron and SSC emission.  On the contrary, the EC
radiation pattern of the layer will be {\it less} concentrated along
the forward direction (with respect to its synchrotron and SSC)
because, in the comoving frame, the layer emits more EC power in the
backward direction (e.g. back to the black hole).  The corresponding
transformations are fully described in Paper I, and we briefly
summarise them in the following. We here follow Dermer (1995),
assuming that, due to the strong aberration, in the rest frame of the
spine (layer) all the photons, assumed to be isotropic in the layer
(spine) frame, can be considered as coming from a single direction,
i.e. along (opposite to) the jet axis. This approximation greatly
simplifies the calculations, allowing us not to consider the detail of
the geometry in the transformations.

The beaming of the synchrotron and SSC emission for both regions
follows the usual $\delta_s^{3+\alpha}$ (spine) and
$\delta_l^{3+\alpha}$ (layer) pattern, 
while the patterns of the EC components are described by more complex
shapes.  In particular, the EC from the spine will have a rather
narrow pattern given by $\delta_{s,l}^{4+2\alpha}
\delta_l^{3+\alpha}$, where $\delta_{s,l}$ is the Doppler factor of
the spine as measured in the layer frame.  
The first term
($\delta_{s,l}^{4+2\alpha}$) describes the same pattern as the EC
radiation produced when scattering the seed photons produced by the
broad line regions of powerful blazars, as firstly pointed out by
Dermer (1995).
It can be understood by going
in the reference frame where the seed photons are isotropic (in this
case this means comoving with the layer).  In this frame we see the EC
from the spine boosted by the ``Dermer" pattern
$\delta_{s,l}^{4+2\alpha}$.
%
%
Finally, we go to the observer frame using the standard $\delta_l^{3+\alpha}$ 
term. 
Analogously, the amplification of the EC from the layer is 
$\propto \delta _{l,s}^{4+2\alpha}  \delta_s^{3+\alpha}$. 
%
%
The net result is that the EC emission from the layer is less boosted
in the observer frame than the corresponding EC emission from the spine.

\begin{figure}
\vskip -0.9 true cm
\centerline{ \psfig{file=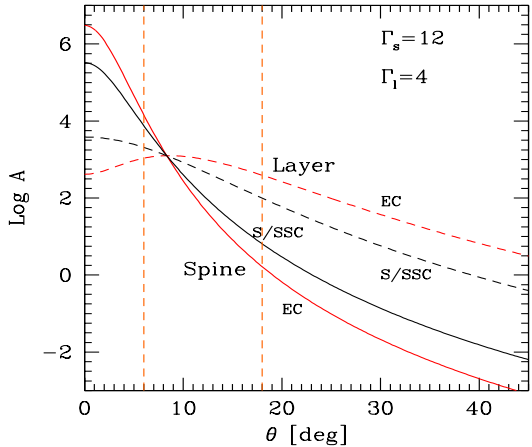,height=8cm,width=9cm}} 
\vskip -1.1 true cm
\caption{
Amplification factors for the different emission components
as a function of the viewing angle. 
Solid (dashed) lines refer to the spine (layer), while black 
(red) lines show the to synchrotron/SSC (EC) factors. 
We assume $\Gamma_s=12$ and $\Gamma_l=4$, the values
we used for the case of M87, and $\alpha =1$. The two vertical
lines indicate the two angles assumed for the calculation of the SEDs,
$6^\circ$ and $18^\circ$. See text for discussion.  
}
\label{fig2}
\end{figure}

Fig. \ref{fig2} illustrates this point, showing the
amplification factors of the different components as a function of the
viewing angle (from 0$^\circ$ to 45$^\circ$), calculated assuming 
$\Gamma_s=12$ and $\Gamma_l=4$ (the values we used for M87, see below).
Solid lines refer to the spine, dashed lines to the layer.  Black and
red lines correspond to the synchrotron/SSC and EC amplification
factors, respectively.  For the spine, the EC amplification
is larger than the Synchrotron/SSC one for small angles, while at
large angles the latter dominates.  
In this sense one can say that the
EC emission has a narrower pattern than the Synchrotron/SSC emission.
The opposite occurs for the layer: at small angles the synchrotron/SSC
amplification is larger then the EC one, which instead becomes more
important at large angles.  
Note, moreover, that the EC/layer emission
has its maximum at $\simeq$10$^\circ$, while all the other curves
peak at 0$^\circ$.  
To understand this, consider the emission pattern in the
frame comoving with the layer.  In this frame the seed photons
produced by the spine are coming along the black hole--layer
direction: as a consequence the relativistic electrons will produce
more EC radiation in the opposite (layer--black hole) direction.  This
is the radiation that will be boosted the least when we transform the
intensity in the observer frame (and the opposite for radiation along
the outward radial direction: in the comoving frame it is the weakest,
but it is boosted the most when going to the observer frame).  This
``compensation" between the intrinsic anisotropy as seen in the
comoving frame and the beaming effect means that the maximum flux will
be observed not at 0$^\circ$, but at a larger angle.  At the extreme,
if the layer is not moving, the maximum will be observed at 180$^\circ$.
This characteristic is particularly relevant for sources
seen at large angles: for those, the EC produced by the layer becomes
relatively more important.

\subsection{Results}
\label{sec3}

We have two possibilities to reproduce the SED with this model: i)
either we assume that the low energy bump is produced by the layer and
the TeV emission originates in the spine or ii) the layer produces the
TeV radiation and the spine mainly contributes at low frequencies.
The first choice is disfavoured in the present case. 
In fact, to produce TeV photons, the typical Lorentz factor of the electrons in
the spine must be large ($\gamma\sim 10^6$) and, assuming typical
magnetic fields of order 0.1--1 G, the corresponding synchrotron
emission would peak in the hard X--ray band. 
Under this conditions the SSC emission would be deeply in the 
Klein--Nishina regime and thus strongly depressed and the TeV 
emission would be dominated by the EC component. 
If the spine SED is dominated by the EC component at large
(18$^\circ$) angles, it will be even more so at small angles (see
Fig. 2), with $L_{\rm TeV}/L_{\rm X}\gg 20$.  
Instead, in TeV BL Lacs, the synchrotron and the IC components show 
comparable luminosities [with few exception, most notably 1ES 1426+428 
(Aharonian et al. 2002), although the TeV luminosity depends on 
the still unclear amount of intergalactic absorption]. 
On the other hand, as we show below, the choice ii) leads to a predicted 
SED for the beamed counterpart of M87 closely resembling those observed in 
LBL sources and thus we decided to adopt this view in the modelling.

\begin{table*} 
\begin{center}
\begin{tabular}{|l|llllllllllll|}
\hline
& $R$    
& $H$  
& $L_{\rm syn}$  
& $B$  
& $\gamma_{\rm min} $  
& $\gamma_{\rm b} $ 
& $\gamma_{\rm cut}$  
& $n_1$
& $n_2$
& $\Gamma$ 
& $\theta$ \\
&cm  &cm &erg s$^{-1}$  &G & & & & & & &deg.  \\
\hline  
spine low  &7.5e15 &3e15 &4.0e41 &1    &600 &5e3 &1e8 &2 &3.7 &12  &18 \\ 
layer low  &7.5e15 &6e16 &1.6e38 &0.35 &1   &2e6 &1e9 &2 &3.7 &4  &18  \\
\hline
spine high &7.5e15 &3e15 &5.2e41 &1    &600 &5e3 &1e8 &2 &3.7 &12  &18 \\ 
layer high &7.5e15 &6e16 &4.0e38 &0.2  &1   &6e6 &1e9 &2 &3.7 &4  &18  \\
\hline
\end{tabular}                                                         
\caption{Input parameters of the models for the layer and the spine
shown in Fig. \ref{fig1}. All quantities (except
the bulk Lorentz factors $\Gamma$ and the viewing angle $\theta$) are
measured in the rest frame of the emitting plasma. 
The external
radius of the layer is fixed to the value $R_2=1.2 \times R$. The
corresponding values of the electron density are: 108.3
cm$^{-3}$ and 140.8 cm$^{-3}$ (spine low and high); 0.74 cm$^{-3}$
and 2.04 cm$^{-3}$ (layer low and high).}
\end{center}
\label{tab1}
\end{table*}                                                                  

The SEDs obtained for the core of M87 with the model are shown in
Fig. \ref{fig1} for both 2004 and 2005 and the parameters are reported
in Table 1. 
 
The model shown is intended as an example of the possible parameters
giving a reasonable shape for the SED.  A whole family of acceptable
solutions, each one with different values for the parameters, is
allowed.  However, the joint request to reproduce the data and to have
a spine SED resembling that of a BL Lac at small angles provides some
constraints.  Since the layer has to produce TeV radiation, the
Lorentz factor of the electrons at the peak is constrained to be of
the order of $\gamma_b \approx 10^6$, independently on the component
(SSC or EC) dominating the high--energy emission.  This, in turn,
fixes the peak of the synchrotron component close to or within the
X--ray band for the chosen magnetic field.  Smaller $B$--values are
allowed, but not larger, since in this case the layer synchrotron
emission would overproduce the observed data.
For the spine, as already discussed, most of the constraints come
from the request that the Lorentz factor, the size and the magnetic
field are similar to those typically inferred for (LBL) blazars.

The model slightly overproduces the emission in the optical
band. However, since the optical and the X--ray emission are clearly
correlated (Perlman et al. 2003) we expect that high X-ray fluxes
correspond to high optical states. Note also that we predict quite
bright GeV emission in this period, exceeding the {\it EGRET} upper
limit by a factor of 3 and easily detectable with {\it GLAST}. The
high-energy component of the layer is dominated by the EC emission
from the beamed (in the layer frame) IR-optical photons from the
spine. On the other hand, since the synchrotron component of the layer
peaks in the X--rays, the EC emission of the spine is strongly
suppressed by the effect of the Klein--Nishina cross section and hence
the high--energy component is mainly SSC. The variability of the TeV
flux from 2004 to 2005 is reproduced by increasing the luminosity
radiated by the layer (and thus the corresponding density of the
electrons) and slightly decreasing the magnetic field. Note also
that the small size of the emitting region allows the rapid changes
such as those detected by H.E.S.S. ($\Delta t\sim 2$ days).

A supplementary contribution to the IC emission from both the spine
and the layer could come from the scattering of the external IR
radiation field, possibly associated to the putative low--efficiency
accretion flow.  The luminosity of the core of M87 in the infrared
band is $L\approx 10^{41}$ erg s$^{-1}$ (Perlman et al. 2001).  This
is an upper limit to the luminosity of the accretion flow, since the
observed power law spectrum supports the idea that it is mainly
associated to the emission from the jet (Perlman et al. 2001).  We can
set an upper limit to the importance of this component by assuming
$L=10^{41}$ erg s$^{-1}$ for the accretion luminosity, produced in a
region whose size is of the order the distance of the active region of
the jet (around $\sim 10^{17}$ cm).  This maximizes the importance of
this external component: for smaller radii the accretion IR radiation
would be strongly deboosted in the jet frame, while for larger radii
it will be more diluted.  Even so, this external component has a
radiation energy density smaller (by more than an order of magnitude)
of the one already present in the spine and in the layer.


Inspection of Fig. \ref{fig1} reveals that we nicely reproduce the
slope of the 2004 TeV spectrum, while the 2005 spectrum appears to be
harder than the model (although the relatively large error bars
prevent a firm conclusion).  The steep slope of the model continuum in
this band is the result of the $\gamma\gamma$ absorption of TeV photons
interacting with optical--IR photons of the spine and thus cannot be
simply avoided.

In Fig. \ref{fig1} we also report (black line) the emission from the
spine relative to 2004 as would appear to an observer almost aligned
($\theta =6^\circ$) with the jet. For comparison we report (orange,
filled symbols) the multifrequency SED for BL Lac itself (the
prototype of BL Lac objects), collected during different campaigns
(see Ravasio et al. 2002 for references) and reporting also the TeV
spectrum recently detected by MAGIC (Albert et al. 2007).

\section{Discussion}
\label{sec4}

We propose that the TeV emission detected from M87 originates in the
structured jet at blazar scale. The most likely possibility is that
the emission from the spine reproduces the low energy component, from
radio to X--rays, while the layer contributes to the bulk of the TeV
radiation. With this choice, the beamed counterpart of M87 observed at
small angle would have a SED closely resembling those of Low--Peaked
BL Lac objects. Correspondingly the spine is characterised by physical
parameters close to those usually inferred for those sources ($\gamma
_b\sim 10^3$, $B\sim 1$ G). The layer, instead, would be characterised
by a low magnetic field ($B\sim 0.1$ G) and large peak Lorentz factors
($\gamma _b\sim 10^6$). It is possible that particles in the layer are
energized by turbulent acceleration (e.g. Stawarz \& Ostrowski 2002).
In this case the expected distribution would have a pile-up at the
energy where acceleration and loss processes are in equilibrium
(e.g. Katarzynski et al. 2006). Suggestively, in our conditions
(losses dominated by IC) the expected equilibrium Lorentz factor would
be close to $\gamma \sim 10^6$ for $B\sim 0.1$ G.

Note that our conclusions are rather different from that of
Georganopoulos et al. (2005) who discussed a similar model relying on
the possibility that the jet experiences a strong deceleration at the
blazar scale. In their model the high-energy emission is produced in
the innermost and faster portion of the jet, while the slow external
portions mainly contributes at low frequency. Under these conditions
the beamed counterpart of M87 would be a TeV BL Lac. However, due to
the different beaming patterns or the synchrotron and IC components
(analogously to our case), the resulting spectrum would be
characterised by a strong dominance of the TeV component with respect
to the X--ray one, contrary to what is usually observed in the known
TeV BL Lacs.

In our model the optical and the X--rays are produced by a region
different from that responsible for the TeV emission. Therefore a
strict correlation between low energy and TeV emission is not directly
required (even if it is possible). Analogously, the MeV-GeV emission
should originate in the spine. Therefore we expect that the emission
possibly detected by {\it GLAST} will not exactly follow the TeV
component.

A potential weak point of our model concerns the slope of the TeV
spectrum. The slope is mainly dictated by the absorption of TeV
photons by the dense optical radiation field and, in this sense, it is
rather robust and does not strongly depend on the underlying slope of
the intrinsic TeV spectrum. In particular, hard spectra such as that
recorded in 2005 are quite difficult to achieve in this scheme. A
possibility to avoid important absorption of gamma-rays would be to
enlarge the emission region, hence diluting the target photon density
and decreasing the absorption optical depth. However the increase of
the source size is limited by the observed short variability
timescales. Further observations with increased sensitivity could
provide stringent constraints to our interpretation.

\section*{Acknowledgements}
We thank M. Chiaberge and B. Balmaverde for useful discussions on the
X-ray data of M87 and the referee, Mark Birkinshaw, for constructive
comments that help to clarify the text.

\end{document}